\documentclass[prb,twocolumn,superscriptaddress]{revtex4}
 \usepackage{amsmath,amssymb,graphicx}
 \DeclareMathOperator{\erf}{erf}

\begin{document}

\title{Multiple defect model for non-monotonic structure relaxation \\ in
binary systems like P\lowercase{d}--E\lowercase{r} alloys charged
with hydrogen}

\author{Albert A. Katsnelson}
\email[Electronic address: ]{albert@solst.phys.msu.su}
\author{Anton Yu. Lavrenov}
\affiliation{Faculty of Physics, M.~V.~Lomonosov Moscow State University,
Moscow 119992, Russia}
\author{Ihor A. Lubashevsky}
 \affiliation{Faculty of Physics, M.~V.~Lomonosov Moscow State University,
Moscow 119992, Russia}
 \affiliation{General Physics Institute, Russian Academy of Sciences, Vavilov
street 38, Moscow 119991, Russia}

\date{\today }

\begin{abstract}

In binary metallic systems like the Pd--Er alloys charged with
hydrogen the observed structure evolution exhibits complex
dynamics. It is characterized by non-monotonic time variations in
an Er-rich fraction respect with an Er-poor fraction observed
experimentally. The present paper proposes a qualitative model for
this non-monotonic structure relaxation. We assume that the alloy
have crystalline defects which trap (or emit) an additional amount
of Er atoms. Hydrogen atoms into the alloy disturb the phase
equilibrium as well as change the defects capacity with respect to
Er atoms. Both of these factors lead to the spatial redistribution
of Er atoms and cause the interface between the Er-rich and the
Er-poor phase to move. The competition of diffusion fluxes in
system is responsible for non-monotonic time variations, for
example, in the relative volume of the enriched phase. We have
found the conditions when the interface motion can change its
direction several times during the system relaxation to a new
equilibrium state. From our point of view this effect is the
essence of the hydrogen induced non-monotonic relaxation observed
in such systems. The numerical simulation confirms the basic
assumptions.
\end{abstract}
 \maketitle

\section{Introduction}

For the last years we obtained a series of experimental data
\cite{1,2,3,4,5,6} showing that palladium alloys (Pd--Mo, Pd--Ta,
Pd--Er, etc.) mechanically strained and also charged with hydrogen
exhibit non-monotonic time variations in the phase structure
during relaxation. Using X-ray diffraction techniques we studied
alloys (Pd--Er in this paper) before and after the hydrogen
charging. A palladium bulk was alloyed with Er of 8 atomic per
cent then homogenized at temperature of 900~${}^{\circ}$C during
24 hours and quenched. Further the specimen surface was
mechanically polished, thereby the superficial layer of thickness
of 2--5~$\mu$m became strained. Charging the specimen with
hydrogen was performed electrolytically, in 4~\% NaF solute twice
distilled water under current flux of 80~mA/cm$^2$ during 60~min.
The X-ray diffraction picture was monitored in real time by the
automatic X-ray diffractometer using monochromatic CuK$_{\alpha1}$
irradiation. After charging with hydrogen the palladium specimen
is kept under room conditions.

At the first step we studied the Pd--Er alloy specimen after the
polishing and before the charging \cite{1,2,3}. All the
diffraction lines were of the Lorentz form in which the maximum
location did not exactly correspond the face-centered cubic
lattice but slightly differ to an individual distance for each
diffraction maximum. This points to the presence of anisotropic
elastic strains in the X-ray reflecting superficial layer that
increase the spacing between the lattice planes parallel to the
specimen surface. Also the diffraction maxima were slightly
widened which indicate a large amount of dislocations in the
reflecting layer. An evidence of the two phase state was not
fixed. Peculiarities of the diffraction picture we relate to the
bulk properties of the polycrystalline specimen. The grain
boundaries have a minor effect because of their small thickness.

At the initial stage of relaxation directly after charging the
homogeneous alloy decomposes on two phases and a large amount of
hydrogen forms and restructs crystalline defects. With time
hydrogen atoms degassing the alloy keeping the structure
substantially non-equilibrium. A certain amount of hydrogen
remains inside the system and bounds with Pd and Er atoms and
defects that allow spatial structure to evolve due to the
interaction between the nonequilibrium crystalline defects and the
new phase clusters. Besides, a large amount of the crystalline
defects accelerates diffusion of the solute atoms promoting the
microstructure evolution. For example, the vacancy concentration
can attain 1--20~\% inside metal--palladium alloys after charging
with hydrogen (Pd--M--H systems) \cite{7,8,9,10}. That not only
leads to the hydrogen induced atom migration but also to forming
metal--hydrogen--vacancy phase.

In general the hydrogen charging changes the state of the Pd--Er
alloy dramatically. Key events follow next.

First, embedded hydrogen atoms increase the lattice parameter
substantially. Under relaxation the elastic strain changes from
the stretching to the contraction. Then it grows during the next
two days and then goes down during the further eight days
decreasing for 25~\%. Hereafter, as time went up to one half year,
no strain variations were detected \cite{1,2,3}.

Second, the Lorentz form of diffraction lines now takes
permanently the bimodal form.

Third, under relaxation the profile of the diffraction line
exhibits quasi-periodic variations. Two components of each
diffraction maximum show oscillatory changes with respect to each
other. That correspond to quasi-periodic variations in the
relative volume, for example, of the phase enriched with Er atoms
with respect to Er-poor phase. It should be noted that changing of
the phase partition changes the Er concentration in it (for
example, the growth of Er-poor phase decreases Er concentration in
it) and, as a result, the difference between Er concentrations in
two phases becomes greater.

\begin{figure}
\begin{center}
\includegraphics[scale=1]{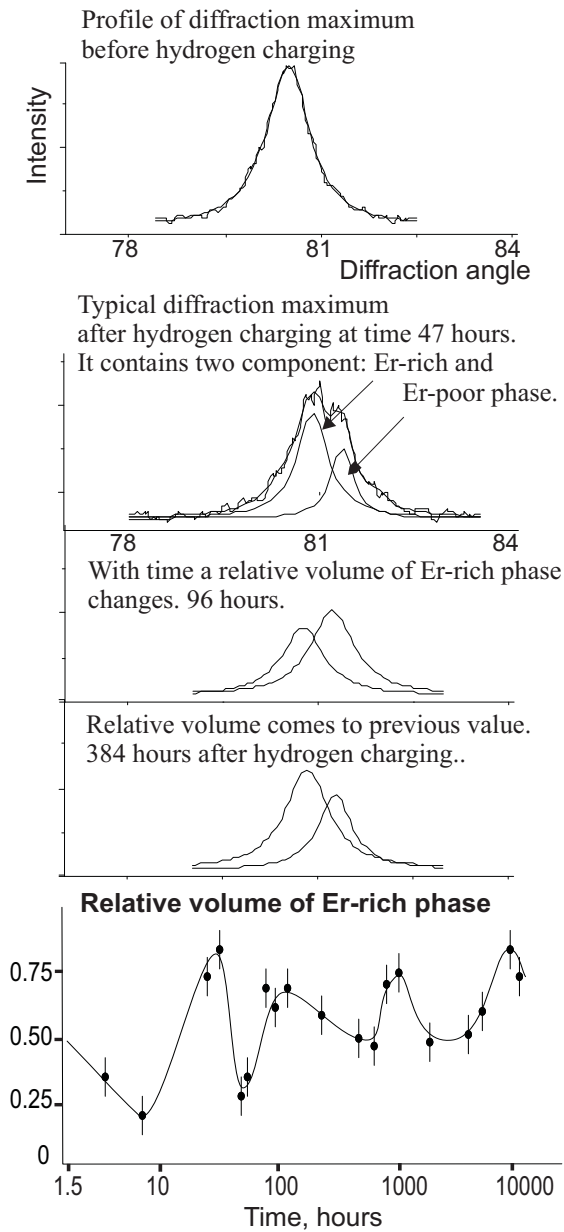}
\end{center}
\caption{Experimental data for the Pd--Er alloy (8 at. \% Er).
\label{f1}}
\end{figure}

The entire phase structure of the Pd--Er alloy at different
Er-concentrations is sufficiently complex \cite{11} and contains
many phases different in chemical composition. However we assume
that corresponding phase transitions did not take place in the
analyzed system because the mole fraction of Er atoms was
sufficiently low, about 10~\%. The observed phase structure of the
Pd--Er alloy evolves the solid solution of Er atoms in the Pd
matrix and an Er-rich phase which should be related in structure
to the ErPd$_7$ intermetallic. Both of them may be modified with
presence of H atoms.

We explain observed phenomena like following. First, the initial
elastic strains are due to defect--metal (D--M) complexes
stretching the crystalline lattice by an image forces. Hydrogen
change the strain sign by the conversion of the stretching D--M
complexes into contracting H--D--M ones. Indeed, due to the high
bond energy of the metal--hydrogen interaction H atom should be
placed inside a defect and then pulls closer metal atoms separated
by the defect. The defect effective volume becomes less than the
volume of the matrix crystal cell.

Second, the affinity of erbium for hydrogen is higher than that of
palladium. Thereby, on one hand, the hydrogen amplifies the Er--Er
interaction. That decreases the stability of the Pd--Er solid
solution and leads to decomposition of homogeneous state on
Er-rich and Er-poor phases. On the other hand, lattice defects
(dislocations, twin boundaries, etc.) contain in nearest volume a
large amount of vacancies. In presence of hydrogen these regions
should attract Er atoms. In other words, the lattice defects begin
to play the role of Er sink. As a result the structure evolution
of the Er--Pd alloy after the hydrogen charging is governed by
several factors which is responsible for the complex behavior of
the system relaxation. Recently Albert Katsnelson \textit{et al.}
\cite{3,4} have proposed a macroscopic phenomenological model for
the non-monotonic relaxation in the framework of synergetic. This
paper presents a microscopic description of studied process.

\section{Model}

What is the essence of the observed phenomenon? We assume that
homogeneous state of the Er--Pd alloy becomes unstable during the
hydrogen charging. A large amount of hydrogen atoms promotes
decomposition on Er-rich and Er--poor phases and also changes the
equilibrium Er-concentrations of both phases. Also Er$_x$H bonds
are originated because of the higher affinity of erbium for
hydrogen with respect to palladium. This process amplifies the
effective Er--Er attraction and changes the phase instability
threshold. That makes nonequilibrium the initially formed phase
regions. The further relaxation is accompanied with the phase
interface motion, leading to variations, for example, in the
relative volume of Er-rich phase. Caused by Er--Er attraction
redistribution of Er atoms takes place first in the close vicinity
of the phase interface $\mathcal{T}$. As a result, near the phase
interface $\mathcal{T}$ the values $c^*_0$ and $c_0$ of the Er
concentration in Er-rich and Er-poor phases attain new equilibrium
values $c^*_0(H)$ and $c_0(H)$ depending certainly on the mean
hydrogen concentration $c_H$ in the Er--Pd alloy at the time of
the experiment. The values $c_0^*(H)$ and $c_0(H)$ do not coincide
with the Er concentrations inside the bulk just after the hydrogen
charging. Thereby Er atoms diffuse away from the phase interface
(or in opposite direction) causing the interface $\mathcal{T}$ to
move. With time the Er distribution near the the interface
$\mathcal{T}$ becomes smoother and the speed of the system
relaxation (interface velocity) becomes slower.

\begin{figure}
\begin{center}
\includegraphics[scale=1]{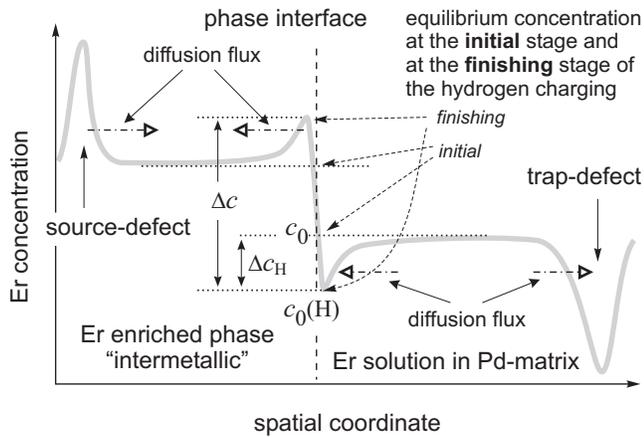}
\end{center}
\caption{Schematic illustration of the Er distribution at the
beginning of the relaxation process. \label{f2}}
\end{figure}

The situation is complicated by crystalline defects
(Fig.~\ref{f2}). On one hand vacancies and theirs complexes in
Er-poor phase plays the role of Er traps. On the other hand
defects inside Er-rich phase can be Er atom sources. Indeed, at
the initial stage of the hydrogen charging defects can attract a
large amount of Er atoms origining probably not only the phase
ErPd$_7$ but also the next phase ErPd$_3$ (see
Ref.~[\onlinecite{11}]). In equilibrium the compound ErPd$_7$ is
intermetallic with the narrow composition interval inside which it
exists. A large value of hydrogen affects the state of the
compound ErPd$_7$ (or rigorously ErPd$_x$H$_y$ where $x\approx7$)
enabling the phase exist inside more wide composition interval. In
this case ErPd$_3$ phase can convert into the main Er-rich phase
(ErPd$_x$H$_y$) releasing extra Er atoms. We see originating of an
Er source. All these defects, traps and sources of Er atoms, give
rise to additional diffusion fluxes competing with the diffusion
fluxes induced by the phase interface. Resulting flux affects the
interface motion. Exactly this flux's competition is responsible
for the non-monotonic relaxation.

\begin{figure}
\begin{center}
\includegraphics[scale=1]{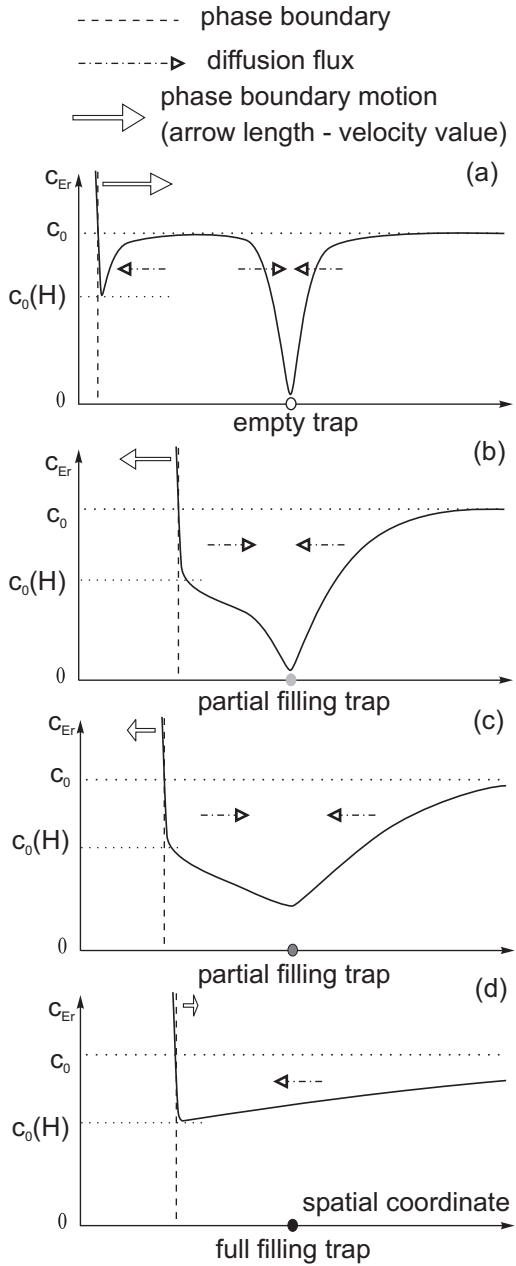}
\end{center}
\caption{Evolution of Er distribution inside Er-poor phase during
the interface motion ($t_a < t_b < t_c <t_d$). \label{f3}}
\end{figure}

Let us discuss this effect in more details (with help of
Fig.~\ref{f2}--\ref{f3}). At the initial stage of the hydrogen
charging the chemical composition of the Er-poor phase (Er solid
solution in Pd-matrix) consists of a higher value of Er
concentration, $c_0$, in comparison with the new equilibrium
concentration $c_0(H)$ at finishing the hydrogen charging. This is
due to H atoms in crystalline lattice amplify the effective Er--Er
attraction. By the same reasons the opposite situation occurs in
the Er-rich phase. In this phase the initial Er concentration is
lower than the equilibrium concentration after the hydrogen
charging. Close by the phase interface $\mathcal{T}$ in the
Er-poor phase the equilibrium values of Er concentration $c_0$ is
constant during the motion of the interface $\mathcal{T}$. It is
due to the fast redistribution of Er atoms in area of the
interface $\mathcal{T}$. The resulting distribution of Er atom and
the induced diffusion flux near the interface $\mathcal{T}$ is
shown in Fig.~\ref{f2}. The velocity $v$ of the interface obeys
the expression
\begin{equation}
 \label{zzz1}
 v \, \Delta c  = D^*\nabla^*_nc - D\nabla_nc\,,
\end{equation}
which depends only on diffusional coefficients $D$ in each of
phases and on gradients of Er-concentration close by the
interface. In this formula the positive direction is shown by the
axis of ordinates in Fig.~\ref{f2}. $\Delta c$ is the drop in Er
concentration $c$ on two sides of the interface $\mathcal{T}$, $D$
and $D^*$ are the diffusion coefficients of Er atoms in Er-poor
phase and Er-rich one, respectively, and, finally, the normal
gradients $\nabla_n^*c$ and $\nabla_nc$ (taken at $\mathcal{T}$)
are directed diffusion fluxes. We relate Er-rich phase to the
ErPd$_7$ intermetallic which can exist only inside a narrow
interval of chemical composition ($\nabla_n^*c$ is small).
Therefore we assume the diffusion flux inside Er-poor phase (Er
solid solution in the Pd-matrix) dominates the interface motion.
In other words, we approximate the rigorous
expression~(\ref{zzz1}) as
\begin{equation}
 \label{zzz2}
 v \,\Delta c \approx D\nabla_nc,,
\end{equation}
which focus our attention on Er-poor phase only. As follow from
expression~(\ref{zzz2}) the phase interface $\mathcal{T}$ should
move inwards Er-poor phase (from left to right in Fig.~2). If this
phase is homogeneous then the interface motion is monotonic
(without changing direction). Its speed decreases with time as the
Er distribution becomes more and more uniform.

The trap-defects in Er-poor phase change the situation
dramatically. Evolution of the Er distribution in the depleted
phase is illustrated in Figs.~\ref{f3}(a)--\ref{f3}(d). Hydrogen
activate some defects as Er atom traps. Near the defects Er
concentration decreases and, as a result, takes a two-minima form
shown in Fig.~\ref{f3}(a). If capacity of the trap is sufficiently
high then with time the Er distribution will take the form shown
in Fig.~\ref{f3}(b). In this case the Er concentration gradient
$\nabla_nc$ at the interface $\mathcal{T}$ changes the sign
causing the interface $\mathcal{T}$ to move in the opposite
direction. As time goes on the defect is filled with Er atoms, the
Er distribution becomes again monotonic and the interface
$\mathcal{T}$ moves in the initial direction. It should be noted
that the source-defects in the Er-rich phase and the trap-defects
in the Er-poor phase affect the interface motion in the same
direction as it follows from Fig.~\ref{f2}. So we will consider
only Er-poor phase.  Er redistribution inside Er-rich phase cannot
change essentially the results of our model. Its can change only
the numerical results which are not principal for today.

Mentioned above process shows changing direction of the interface
motion two time, or shows one quasi-periodic fragment of the time
variations in the phase partition. At the end of this fragment teh
Er distribution becomes smoother (Fig.~\ref{f3}(d)) and the
interface velocity is less than at beginning. The following
quasi-periodic fragments of the non-monotonic relaxation is due to
other defects in the Pd-matrix. We need to develop a conception of
their activation as traps or sources of Er atoms.

\begin{figure}
\begin{center}
\includegraphics[scale=1]{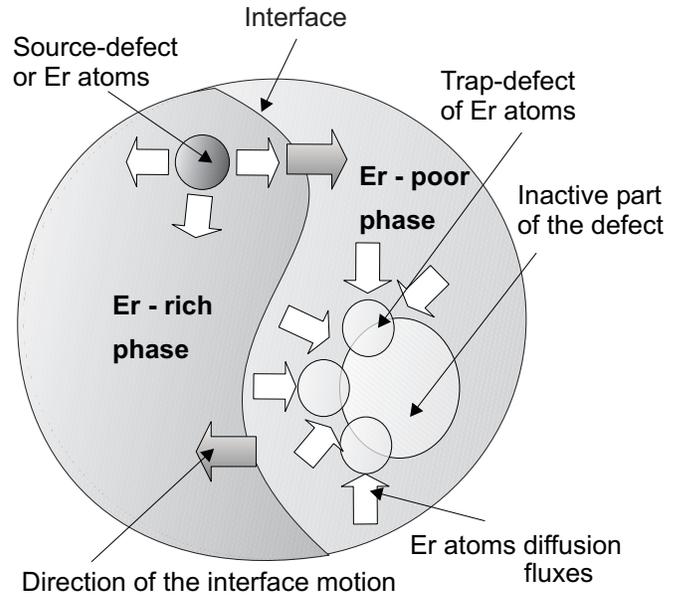}
\end{center}
\caption{Scheme of competing diffusion fluxes inside one domain of
the polycrystal. The process of non-monotonic relaxation is kept
by defects of different types activated at different time after
hydrogen charging. \label{f4}}
\end{figure}

Before the hydrogen charging defects expand Pd-matrix and the
structure of the crystalline defects should reflect this feature.
Conversely after the hydrogen charging defects squeeze the
lattice. Under this conditions reconstruction of the defect should
take a certain time before atoms rearrange themselves. It is quite
reasonable that this process meets a certain potential barrier of
the lattice elasticity nature. The elastic stress caused by other
defects can make this barrier higher. But when nearest defects are
reconstructed completely then the potential barrier is negated
because of atoms arrange its structure. In this case
reconstruction of the given defect becomes more easy. From this
point of view we assume that the defect will be activated as the
trap of Er atom only when the preceding defect is practically
filled with Er atoms. In other words, the defect activation during
the interface motion proceeds like the domino effect. Each act of
the defect activation and filling with Er atoms corresponds to one
quasi-periodic fragment of the non-monotonic relaxation. Besides
there are a number of crystalline defects of different size and
kind \cite{10} which can make the relaxation even more complex.
The larger is a defect, the higher is the potential barrier of
atom rearrangement and, so, the more is the time needed for its
reconstruction. Thereby small defects should be activated first
and only then large defects can come into play (see
Fig.~\ref{f4}). The more is the distance between the defect and
the interface (or the bigger is the defect), the more is the time
of Er redistribution near the interface. Therefore at the further
stage of the relaxation the duration of quasi-periodic fragments
is expanded remarkable.

In the present paper we will consider only characteristics of
defects which are needed us for qualitative analysis. Our aim is
point reader to key moments of structure evolution in Pd--Er--H
system. Quantitative description is the work of future. Now we can
formulate the governing equation simulating the interface motion
and the non-monotonic relaxation.

\section{Governing equations}

Taking into account aforesaid we will consider the diffusion of Er
atoms inside Er-poor phase (regarded as the half-axis $z
> z_{\mathcal{T}}$) bounded by the phase interface $\mathcal{T}$.
We write the diffusion equation for the atomic concentration $c$
of Er atoms in this phase as
\begin{equation}
  \frac{\partial c}{\partial t} = D\frac{\partial^{2}c}{\partial z^{2}}
  -c\frac{l}{\tau_{\mathrm{tr}}} \sideset{}{'}\sum_{i=1}^{\infty}q_iJ_i(q_{i-1})
  \delta (z-z_i)\,.
  \label{1}
\end{equation}
Here $D$ is the diffusion coefficient of Er atoms (which is
constant in this article), the trap-defects are approximated as
the $\delta$-like sinks placed one from other at the equidistant
distance $\{z_i=iz^0_{\text{tr}}\}_{i=1}^\infty$, and the prime
means that the sum runs over all the defects located in the
Er-poor phase, $z_i>z_{\mathcal{T}}$. A $\delta$-sink is
characterized by the real physical size $l$ of a trap-defect, by
the average time $\tau_{\text{tr}}$ of trapping an Er
atom\footnote{The time $\tau_{\text{tr}}$ should not be confused
with the lifetime of Er atoms inside the defect which infinitely
long in the given model.}, and by the defect capacity $q$, i.e.
the number of free seats for Er atoms at the current moment of
time.

At the initial time $t=0$ the capacities of all the defects have
the same value $q_0$. As time goes on the active defect $i$ traps
Er atoms decreasing the capacity $q_i$. We assume defect $i$ to be
activated when the capacity of the preceding defect $i-1$ drops
down the threshold $q_c=\theta_c q_0$ where $\theta_c\ll 1$. This
behaviour is described by the activation function for $i\geq2$
\begin{equation}
 \label{2}
  J_i(q_{i-1})=
 \begin{cases}
    0, & \text{if $q_{i-1} \geq q_c$},\\
    1, & \text{if $q_{i-1} <    q_c$.}
 \end{cases}
\end{equation}
For the first defect, by definition, $J_1=1$, because we assume it
becomes active by the phase interface directly. The defect
capacity obeys the equation
\begin{equation}
  \label{fill}
  \frac{dq_i}{dt} = -\frac{l}{a\tau_{\text{tr}}}\,c(z_i)q_iJ_i(q_{i-1})\,,
\end{equation}
where $a$ is the spacing of the Pd-lattice.

The phase interface $\mathcal{T}$ initially is located at zero
point
\begin{equation}\label{3}
    z_{\mathcal{T}}|_{t=0}=0\,.
\end{equation}
Its further motion is governed by the diffusion flux of Er atoms inwards the
Er-poor phase and obeys the equation (see formula~(\ref{zzz2}))
\begin{equation}\label{4}
  v\stackrel{\text{def}}{=} \frac{dz_{\mathcal{T}}}{dt} =
  \frac{D}{\Delta c}\left.\frac{\partial c}{\partial
  z}\right|_{z_{\mathcal{T}}},
\end{equation}
where the value $\Delta c$ is shown in Fig.~\ref{f2}. In writing
expression~(\ref{4}) we ignored the effect of diffusion flux
inside the Er-rich phase on the interface motion. Besides when the
phase interface crosses the point $z_i$ (so defect $i$ appears in
the Er-rich phase) we assume it not affect the interface motion
further. This is justified by the fact that the interface
$\mathcal{T}$ can pass through the defect location if only it is
filled with Er atoms and, thus, invisible to the interface.

At the interface $\mathcal{T}$ the Er-concentration takes the new
quasi-equilibrium value which formed by fast local Er
redistribution near $\mathcal{T}$ (Fig.~\ref{f2}):
\begin{equation}\label{5}
  c(z_{\mathcal{T}})= c_0 - \Delta c_{\mathrm{H}}\,,
\end{equation}
where $c_0$ is the initial Er-concentration inside the Pd-matrix, i.e.
\begin{equation}\label{6}
  c(z,t)|_{t=0} = c_0 \quad\text{for $z>0.$}
\end{equation}

The system of equations~(\ref{1}), (\ref{fill}), (\ref{4}) with
the boundary condition~(\ref{5}) and the initial
conditions~(\ref{3}), (\ref{6}) forms our model of the phase
partition evolution or, generally speaking, the non-monotonic
relaxation.

\subsection{One quasi-periodic fragment of non-monotonic relaxation}

Let us, first, find the main features of the model and write
parameters (in dimentionless form) governing the structure
relaxation. To do this we consider the beginning of the interface
motion when the only one nearest defect is active. In the next
section we will study the array of defects.

For convenience we attach our frame of reference to the moving
interface. We will use in formula dimensionless time $\tau =
(Dt)/(z^0_{\text{tr}})^2$, coordinate
$\xi=(z-z_{\mathcal{T}})/z^0_{\text{tr}}$, normalized Er
concentration $\eta = c/c_0$, the defect capacity $\theta =
q_1/q_0$ and the dimensionless time-dependent coordinate
$\xi_{\text{tr}}(\tau) =
(z^0_{\text{tr}}-z_{\mathcal{T}}(t))/z^0_{\text{tr}}$ of the first
defect in the moving frame. In these terms the aforestated system
is converted to the following equations:
\begin{eqnarray}
 \label{7}
 \frac{\partial \eta }{\partial \tau }+\frac{d\xi _{\text{tr}}(\tau )}{d\tau }
 \frac{\partial \eta }{\partial \xi } &=&\frac{\partial ^{2}\eta }{\partial
 \xi ^{2}}-\Omega \eta \theta \delta \left[\xi -\xi _{\text{tr}}(\tau )\right], \\
 \label{8}
 \frac{d\theta }{d\tau } &=&-\Lambda \Omega \eta
 (\xi_{\text{tr}},\tau)\theta.
\end{eqnarray}
Equations~(\ref{7}), (\ref{8}) must be completed by the boundary
conditions for $\tau>0$:
\begin{eqnarray}
 \label{9a}
 \left. \frac{\partial \eta (\xi ,\tau )}{\partial \xi }\right| _{\xi =0}
 & = &
 -\frac{\Delta c}{c_{0}}\,\frac{d\xi_{\text{tr}}(\tau )}{d\tau },\\
 \label{9b}
 \left.\eta (\xi ,\tau )\right| _{\xi =0}
 & = &
 1-\frac{\Delta c_{\mathrm{H}}}{c_{0}}
\end{eqnarray}
and the initial conditions:
\begin{equation}
 \label{10}
 \left.\eta(\xi ,\tau )\right|_{\tau =0}=1\,,\quad
 \left.\theta(\tau )\right|_{\tau =0}=1\,,\quad
 \xi_{\text{tr}}(\tau )|_{\tau =0}=1
\end{equation}
for $\xi > 0$. Following quantities:
\begin{equation}
 \label{11}
 \frac{\Delta c}{c_{0}},\quad \frac{\Delta c_{\mathrm{H}}}{c_{0}},\quad
 \Omega =\frac{q_{0}lz_{\text{tr}}^{0}}{D\tau _{\mathrm{tr}}},\quad \Lambda =
 \frac{c_{0}z_{\text{tr}}^{0}}{q_{0}a}
\end{equation}
form the dimensionless parameters governing the model.

$\Omega$ is one of the key parameters of the model. If $\Omega\ll
1$ (which corresponds to small capacity of the defect $q_0$ or to
very fast diffusion) then the trap-defect not affect the interface
motion. Lets $\Omega$ to be of order unity. At the initial stage
when $\tau \ll 1$ (i.e. $t\ll (z^0_{\text{tr}})^2/D$) the only Er
redistribution near the interface $\mathcal{T}$ governs its
dynamics. In this case the latter term on the right-hand side of
Eq.~(\ref{7}) can be ignored and the system of equation~(\ref{7}),
(\ref{9a})--(\ref{10}) has a self-similar solution:
\begin{equation}
 \label{12}
 \eta^{\ast}(\xi,\tau) = 1-\sqrt{\pi}\,\frac{\Delta c}{c_{0}}\,\vartheta_{0}
 e^{\vartheta_0^2}\left[1-\erf\left(\frac{\xi}
 {2\sqrt{\tau }}+\vartheta_{0}\right) \right] ,
\end{equation}
where $\vartheta_{0}$ is the root of the transcendental equation:
\begin{equation}
 \label{13}
 \sqrt{\pi}\,
 \vartheta_{0}e^{\vartheta_0^2}\left[1-\erf\left(
 \vartheta_{0}\right)\right] = \frac{\Delta c_{\text{H}}}{\Delta c}\,.
\end{equation}
The corresponding value of the dimensionless interface velocity
$\vartheta = vz^0_{\text{tr}}/D$ is written as
$\vartheta=\vartheta_{0}/\sqrt{\tau}$. For numerical experiment we
take next values: $\Delta c/c_{0}=1.5$ and $\Delta
c_{\text{H}}/c_{0}=0.5$. In this case the value of $\vartheta_0$
is approximately equal to $\vartheta_{0}\approx 0.2$. For the
given values the interface should reach the first defect (with
$\Omega \ll 1$) in time $\tau\approx 10$.

Now let us activate the defect with $\Omega = 1$. As follows from
Eq.~(\ref{7}) for a sufficiently short time $\tau\sim \Omega^{-2}$
all the Er atoms near the trap-defect will be trapped. Therefore
in the vicinity of the defect Er distribution takes the form shown
in Fig.~\ref{f3}(a). The general form of Er distribution near the
defect remains unchanged until either the interface $\mathcal{T}$
comes close to the defect or the defect is filled with Er atoms.
Certainly, the distribution width will increase with time as
$\sqrt{\tau}$. From  Eq.~(\ref{7}) for $\Omega^{-2}\ll \tau \ll 1$
Er distribution near the defect can be approximated as
\begin{equation}
  \label{14}
  \eta
  (\xi,\tau)=\erf\left(\frac{\left|\xi-\xi_{\text{tr}}\right|}
  {2\sqrt{\tau }}\right).
\end{equation}
Within the same time interval the dimensionless rate of trapping
Er atoms by the defect is estimated as
\begin{multline}
 \label{15}
 \int d\xi \,\Omega\theta\eta\delta(\xi-\xi_{\text{tr}})\\
 \approx \Omega \eta
 (\xi_{\text{tr}},\tau )
  \approx 2\left.\frac{\partial\eta}{\partial\xi}\right|
 _{\xi =\xi_{\text{tr}}+0}\approx \frac{2}{\sqrt{\pi\tau}}\,.
\end{multline}
Because of diffusion dithering at time $\tau \sim 1$ Er
distribution will get the form shown in Fig.~\ref{f3}(b). Such a
distribution has a negative value of the boundary gradient, so the
interface $\mathcal{T}$ moves in the opposite direction (according
to expression~(\ref{9a}) or (\ref{4})). As time goes on further
the defect will be filled with Er atoms. The Er concentration in
area of the defect will grow and the defect will affect weaker the
interface motion. The corresponding form of Er distribution is
illustrated in Fig.~\ref{f3}(c). As follows from Eq.~(\ref{8}) and
estimate~(\ref{15}) the time of the defect filling is about $\tau
\sim \Lambda^{-2}$. After that the defect cannot affect the
interface motion. Er distribution tends again to the self-similar
form~(\ref{12}). For observer the interface velocity changes the
sing for the second time, and the interface motion returns to the
initial direction until the next trap-defect is activated.

\begin{figure}
\begin{center}
\includegraphics[scale=1]{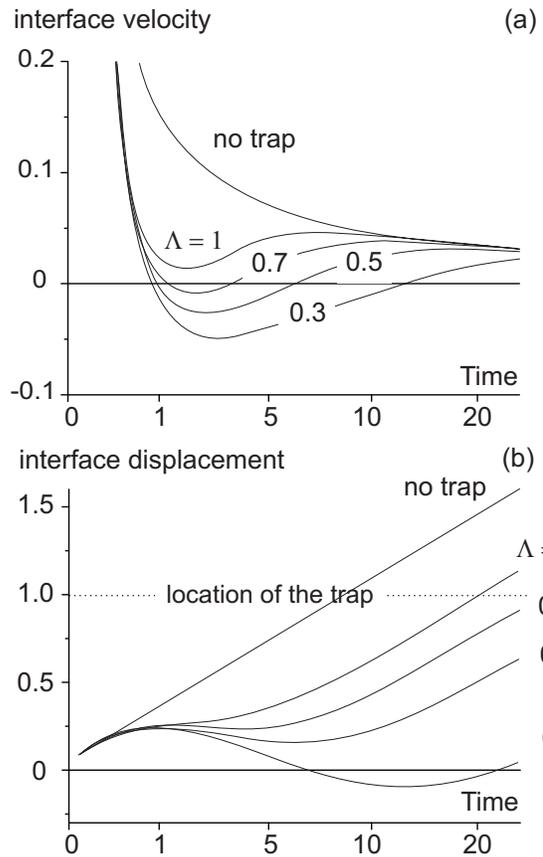}
\end{center}
\caption{The interface dynamics described by the
system~(\ref{7})--(\ref{10}) for different values of the defect
capacity. The fragment ($a$) shows the dimensionless interface
velocity \textit{vs} time and the fragment ($b$) illustrates the
time dependence of the interface displacement. The abscissa
exhibits time at root-square-law scale. In numerical simulation we
used: $\Delta c /c_{0}=1.5$, $\Delta c_{\text{H}}/c_{0}=0.5$, and
$\Omega = 1$. The values of the parameter $\Lambda$ are pointed
out at the curves.\label{f5}}
\end{figure}

The aforementioned scenario describes one quasi-periodic fragment
of the non-monotonic structure relaxation. For numerical analysis
of~(\ref{7})--(\ref{10}) we take $\Delta c /c_{0}=1.5$, $\Delta
c_{\text{H}}/c_{0}=0.5$, $\Omega = 1$ and then vary parameter
$\Lambda$ (in non-mathematical words we study defects of different
capacities). The results are shown in Fig.~\ref{f5}. When the
capacity of the defect is sufficiently high ($\Lambda \le 1$) we
obtain an essential backward interface motion. Otherwise, the
trap-defect will be filled before the induced diffusion flux could
affect the interface motion.

When the first defect is practically filled with Er-atoms the next
defect will be activated. That should give rise to a similar one
quasi-periodic fragment of the non-monotonic relaxation, which is
the subject of the next section.

\subsection{Multi defect effect on the interface dynamics}

In this section we will use the same dimensionless time $\tau$,
coordinate $\xi$, the normalized Er concentration $\eta$ and
defect capacities $\{\theta_i=q_i/q_0\}$. In these terms the
initial system of governing equations is reduced, first, to the
diffusion equation written in the form
\begin{gather}
 \label{16}
 \frac{\partial \eta }{\partial \tau }-\vartheta
 \frac{\partial \eta }{\partial \xi } =
 \frac{\partial ^{2}\eta }{\partial\xi ^{2}}-
 \Omega\eta \sideset{}{'}\sum_{i=1}^\infty\theta_i J_i(\theta_{i-1})
 \delta \left[\xi -\xi_i(\tau )\right],\\
\intertext{where according to (\ref{2}) for $i\geq 2$}
 \nonumber
 J_i(\theta_{i-1})=
 \begin{cases}
    0, & \text{if $\theta_{i-1} \geq \theta_c$},\\
    1, & \text{if $\theta_{i-1} <    \theta_c$,}
 \end{cases}
 \quad \text{and}\quad J_1 = 1.
\end{gather}
In expression (\ref{16}) the value $\vartheta=vz^0_{\text{tr}}/D$,
as before, is the dimensionless velocity of the interface
$\mathcal{T}$ and the sum runs over all the defects located inside
the Er-poor phase (whose coordinates $\xi_i(\tau)>0$ are positive
in the frame attached to the moving interface $\mathcal{T}$). By
definition $\xi_i(\tau) = i -
z_{\mathcal{T}}(\tau)/z^0_{\text{tr}}$ all the quantities
$\xi_i(\tau)$ changes with time at the same rate
\begin{equation}\label{17}
  \frac{d\xi_i(\tau)}{d\tau} = -\vartheta\,.
\end{equation}
Second, the defects filling is described by the system of
equations
\begin{equation}\label{18}
 \frac{d\theta_i}{d\tau} = -\Lambda
 \Omega\eta(\xi_i(\tau),\tau)\theta_iJ_i(\theta_{i-1})\,.
\end{equation}
Third, the interface velocity $\vartheta$ relates with the
boundary condition as~(\ref{9a}):
\begin{equation}\label{19}
  \left. \frac{\partial \eta (\xi ,\tau )}{\partial \xi }\right| _{\xi =0}
  =
  \frac{\Delta c}{c_{0}}\,\vartheta\,.
\end{equation}
The boundary condition (\ref{9b}) holds. The initial
condition~(\ref{10}) should be replaced by the following one
\begin{equation}
 \label{20}
 \left.\eta(\xi ,\tau )\right|_{\tau =0}=1\,,\quad
 \left.\theta_i(\tau )\right|_{\tau =0}=1\,,\quad
 \xi_i(\tau )|_{\tau =0}=i
\end{equation}
for $\xi>0$.

\begin{figure}
\begin{center}
\includegraphics[scale=1]{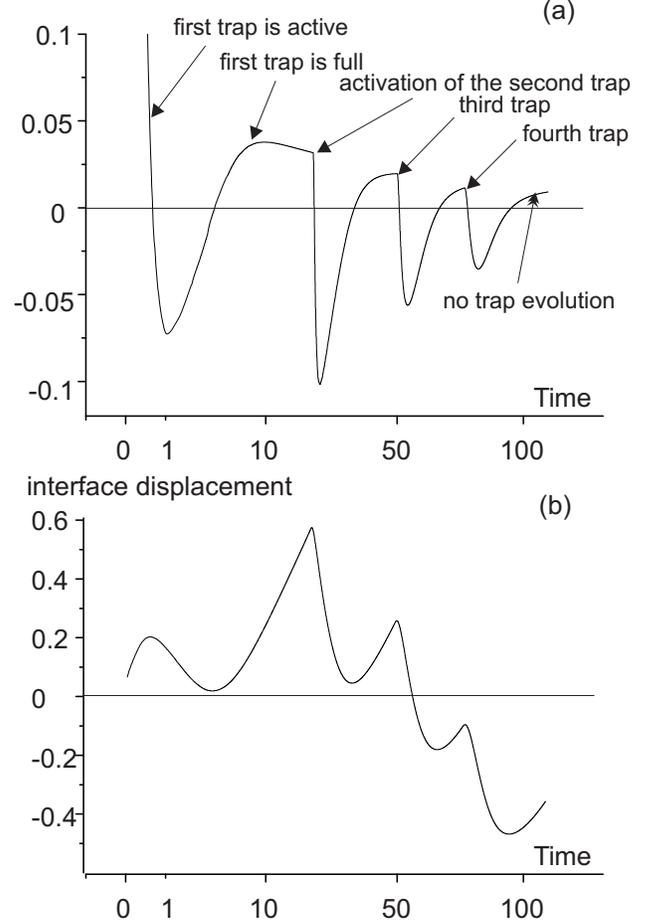}
\end{center}
\caption{Dynamics of the interface $\mathcal{T}$ described by the
system~(\ref{16})--(\ref{20}) in which the interface motion is
affected by four trap-defects activated sequentially. The fragment
($a$) shows the dimensionless interface velocity \textit{vs} time
and the fragment ($b$) illustrates the corresponding time
dependence of the interface displacement. The abscissa exhibits
time at root-square-law scale. In numerical simulation we used the
following values of parameters $\Delta c /c_{0}=1.5$, $\Delta
c_{\text{H}}/c_{0}=0.5$, $\Lambda=0.5$, $\theta_c=10^{-4}$, and
set $\Omega=2$. \label{f6}}
\end{figure}

\begin{figure}
\begin{center}
\includegraphics[scale=1]{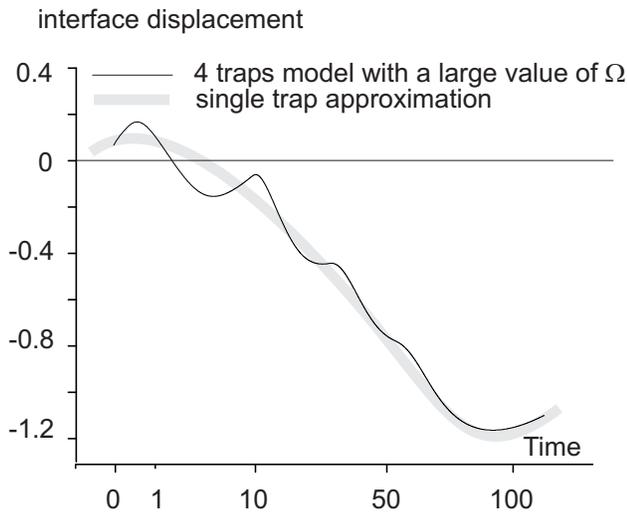}
\end{center}
\caption{Dynamics of the interface when the Er diffusion is fast.
It show the interface displacement \textit{vs} time predicted by
the system~(\ref{16})--(\ref{20}) when the interface motion is
affected by four trap-defects activated sequentially. The abscissa
exhibits time at root-square-law scale. In numerical simulation we
used the following values of parameters $\Delta c /c_{0}=1.5$,
$\Delta c_{\text{H}}/c_{0}=0.5$, $\Lambda=0.5$,
$\theta_c=10^{-4}$, and set $\Omega=10$. \label{f7}}
\end{figure}

The system of equations has been analyzed numerically for the same
values of the parameters $\Delta c /c_{0}=1.5$, $\Delta
c_{\text{H}}/c_{0}=0.5$ but with fixed the value of the parameter
$\Lambda$ equal to $\Lambda=0.5$. In other words here we fix the
defect capacity and analyze the interface dynamics depending on Er
diffusivity and the threshold value $\theta_c$. At first we set
$\theta_c=10^{-4}$ and $\Omega=2$. The corresponding dynamics of
the interface $\mathcal{T}$ is shown in Fig.~\ref{f6}. Similar
non-monotonic structure evolution is observed for $\Omega\sim 1$.
When the parameter $\Omega$ takes sufficiently large values (i.e.
if the Er diffusivity $D$ is small enough) so that $\Omega\gg 1$
then the trap-defects affect the interface motion as one trap but
of much higher capacity. For example, Fig.~\ref{f7} exhibits the
interface dynamics for $\Omega=10$. In the given case the
interface motion looks like one quasi-periodic fragment but
prolonged substantially. For $\Omega\ll 1$, i.e. when the Er
diffusion is sufficiently fast the defects will be filled also
fast and they have no considerable effect on the interface motion.

The specific details of the interface dynamics depend on the
threshold $\theta_c$ of the defect activation. In particular, for
$\theta_c=10^{-2}$ we observe the interface dynamics similar to
Fig.~\ref{f6} for $\Omega=1$ whereas the value $\Omega=2$ got us
to the interface motion of the same form as shown in
Fig.~\ref{f7}.

It should be noted that in Fig.~\ref{f5}--\ref{f7} the abscissa is
a time at root-square-law scale in order to emphasize the
oscillatory behaviour of the non-monotonic structure relaxation.
In reality each following quasi-periodic fragment is more
prolonged than the preceding one. A similar behaviour was found
experimentally (see Fig.~\ref{f1}). However, the experimental
curve exhibits quasi-periodic time variations at logarithmic scale
rather than root-square scale for the theoretical curve. We
explain this discrepancy by taking into account different size of
defects. When the larger a defect is, the later it comes into play
and the longer time it will be filled with Er.

\section{Conclusion}

We have proposed a mechanism which responsible for the
nonmonotonic structure relaxation observed in metallic binary
systems like the Pd--Er alloys after hydrogen charging. The
observed quasi-periodic alterations of the diffraction line
profile are related with time variations of volume and
concentration of two phase (Er-rich and Er-poor) arose in the
alloy after hydrogen charging. We developed the model in which we
consider active crystalline defects that trap additional amount of
Er atoms. The key point of the model is disturbance by hydrogen
the phase equilibrium as well as change the defect capacity with
respect to Er atoms. Both of these factors lead to the spatial
redistribution of Er atoms causing the interface to move between
the two phases. Because of the induced diffusion fluxes these
factors individually would move the interface in the opposite
directions. As a result their competition is responsible for
non-monotonic time variations, for example, in the relative volume
of Er-rich phase observed experimentally.

We have found the conditions when the interface motion can change its
direction several times during the system relaxation to a new equilibrium
state. The latter effect is, from our point of view, the essence of the
hydrogen induced non-monotonic relaxation observed in such systems. The
numerical simulation confirms the basic assumptions.

It should be also noted that the present paper pretends only to a
qualitative description of the observed non-monotonic structure
relaxation. Actually we have singled out the basic feature of such
systems that is responsible for this complex process. To compare
the experimental data with results of numerical simulation a more
sophisticated model should be developed.

\begin{acknowledgments}
  This work was supported by the Russian Foundation of Basic Research,
  grants~\#~02-02-16537, \#~01-01-00389 and \#~02-02-06167, and by the
  INTAS, grant~\#~00-0847.
\end{acknowledgments}

\end{document}